# *Hubble Space Telescope* Imaging of the

# Ultracompact Blue Dwarf Galaxy HS 0822+3542:

# An Assembling Galaxy in a Local Void?

Michael R. Corbin[1], William D. Vacca[2], John E. Hibbard[3],

Rachel S. Somerville[4]  & Rogier A. Windhorst[1]

---

[1]Department of Physics & Astronomy, Arizona State University, P.O. Box 871504, Tempe, AZ  85287; Michael.Corbin@asu.edu, Rogier.Windhorst@asu.edu

[2]USRA, MS 144-2, NASA-Ames Research Center, Mountain View, CA, 94025; wvacca@mail.arc.nasa.gov

[3]National Radio Astronomy Observatory, 520 Edgemont Rd., Charlottesville, VA 22903; jhibbard@nrao.edu

[4]Space Telescope Science Institute, 3700 San Martin Dr., Baltimore, MD, 21218; somerville@stsci.edu

ABSTRACT

We present deep $U$, narrow-$V$, and $I$ band images and photometry of the ultracompact blue dwarf galaxy HS 0822+3542 obtained with the Advanced Camera for Surveys / High Resolution Channel of the *Hubble Space Telescope.* This object is also extremely metal-poor (12 + log (O/H) = 7.45) and resides in a nearby void. The images resolve it into two physically separate components that were previously described as star clusters in a single galaxy. The primary component is only ~100 pc in maximum extent, and consists of a starburst region surrounded by a ring-like structure of relatively redder stars. The secondary component is ~50 pc in size and lies at a projected distance of ~80 pc away from the primary, and is also actively star-forming. We estimate masses ~$10^7$ M$_\odot$ and ~$10^6$ M$_\odot$ for the two components based on their luminosities, with an associated dynamical timescale for the system of a few Myr. This timescale and the structure of the components suggests that a collision between them triggered their starbursts. The spectral energy distributions of both components can be fitted by the combination of a recent (few Myr old) starburst and an evolved (several Gyr old) underlying stellar population, similar to larger blue compact dwarf galaxies. This indicates that despite its metal deficiency the object is not forming its first generation of stars. However, the small sizes and masses of the two components suggests that HS 0822+3542 represents a dwarf galaxy in the process of assembling from clumps of stars intermediate in size between globular clusters and objects previously classified as galaxies. Its relatively high ratio of neutral gas mass to stellar mass (~1) and high specific star-formation rate, log(SFR/M$_\odot$) $\cong$ -9.2, also suggest that it is still converting much of its gas to stars.

*Subject Headings*:  galaxies:indvidual (HS 0822+3542) – galaxies: stellar content – galaxies: formation – galaxies:dwarf – galaxies: starburst

## 1. INTRODUCTION

The blue compact dwarf galaxy HS 0822+3542 (= SDSS J082555.44+353231.9) is one of the most metal-poor galaxies known, having 12 + log(O/H) = 7.45±0.02 as measured from its strong nebular emission lines (Kniazev et al. 2000; Kniazev et al. 2003). Its redshift of $z$ = 0.00233 (Kniazev et al. 2003) corresponds to a distance of only 12.7 Mpc under a cosmology of $H_0$ = 71 km s$^{-1}$ Mpc$^{-1}$, $\Omega_\Lambda$ = 0.73, $\Omega_M$ = 0.27, after correction to the frame of the cosmic microwave background. It is however very faint and blue, with a Sloan Digital Sky Survey[a] $g$ magnitude of 17.64 ($M_g$ = -12.88) and $g$ - $r$ = -0.08. HS 0822+3542 is also "ultracompact" under the criteria of having an angular size of just a few arcseconds as measured from ground-based images (Kniazev et al. 2000; Gil de Paz, Madore & Pevunova 2003), and a physical size less than 1 kpc. It resides in a void in the direction of Lynx-Cancer, with the nearest large galaxy more than 3 Mpc away (Pustilnik et al. 2003; hereafter P03). P03 suggest that its star formation has been triggered by interaction with the low-surface brightness dwarf galaxy SAO 0822+3545, located at a projected distance of 3.5′ (13 kpc) away and having a redshift difference of only ~22 km s$^{-1}$.

In this *Letter* we present images and photometry of HS 0822+3542 obtained with the *Hubble Space Telescope* Advanced Camera for Surveys / High-Resolution Channel (ACS/HRC). Our data on HS 0822+3542 are part of a larger program to obtain ACS/HRC images of a sample of ultracompact blue dwarf galaxies (UCBDs) selected from the Sloan Digital Sky Survey. Our primary goal is to resolve the structure of these objects in order to test the hypothesis that they are very small dwarf galaxies in the process of assembling from smaller clumps of gas and stars, as appears to be the case for the prototypical UCBD, POX 186 (Corbin & Vacca 2002). Our results for HS 0822+3542 support this hypothesis, as the ACS/HRC images resolve the galaxy into two physically separate but very small (~100 pc) clumps of stars with complex internal structure that may ultimately coalesce and form a single quiescent dwarf galaxy.

[a] see http://www.sdss.org

These clumps were also detected in the deconvolved ground-based images of P03, but were interpreted as star clusters within a single galaxy. We also find evidence of an underlying old (~ 10 Gyr) stellar population within both clumps of stars, indicating that despite its metal deficiency the object is not forming its first generation of stars.

## 2. OBSERVATIONS AND RESULTS

### 2.1 HST Observations

*HST* ACS/HRC observations of HS 0822+3542 were made on UT 2004 October 4. The filters and total integration times were F330W ( *U*) (1666s), F550M ("narrow *V*") (1906s), and F814W (*I*) (1366s). The limiting $3\sigma$ AB magnitudes for each final image are approximately 25.9 (F330W), 25.7 (F550M) and 26.5 (F814W). These filters were chosen to cover a wide wavelength range and to exclude strong emission such as [O II] $\lambda 3727$, [O III] $\lambda\lambda 4959$, 5007 and H$\alpha$. The F814W filter contains the lines He I $\lambda 7065$ and [Ar III] $\lambda 7136$, but from the measured fluxes of these lines in P03 we estimate their contribution to the total flux measured in the filter to be < 1%. The images were taken in a two-point dither pattern to optimize cosmic ray rejection, and the sub-images in each filter were combined with the STSDAS task "multidrizzle" to produce final cleaned images at the nominal ACS/HRC resolution of $0.027''$ pix$^{-1}$. We measure the FWHM of the point sources in the images to be approximately 2.8 pixels, corresponding to a resolution of approximately 5 pc at the adopted distance to the galaxy. A correction for Galactic foreground extinction was applied to each image based on the maps of Schlegel, Finkbeiner & Davis (1998). From the Sloan Digital Sky Survey spectrum of the galaxy (see Kniazev et al. 2003) we measure a very small amount of internal extinction, E($B - V$) $\cong 0.04$, indicating little internal dust. We also confirm the Kniazev et al. (2003) metalicity value of $12 + \log(\text{O/H}) = 7.45 \pm 0.02$ to within uncertainties.

Figure 1 presents a color composite of the F330W, F550M and F814W images. It also shows a gray-scale version of this image centered on the central part of the object in order to better display its internal structure.  We confirm the result of P03 that the brightest portion of the object consists of two components, which we designate as A and B (corresponding to their components "a" and "b" ).  We resolve their component "c" into a single star, and their component "d" into two separate stars whose absolute magnitudes (see § 2.2) indicate them to be supergiant stars associated with the galaxy.  Our main finding however is that A and B are not individual star clusters within a larger galaxy, as discussed by P03, but are physically separate objects with complex internal structure.  Component A consists of a central starburst region surrounded by a ring-like structure of relatively redder stars, while component B is more irregular.  The angular sizes of both components are also very small, corresponding to small physical sizes at the adopted distance to the galaxy of 12.7 Mpc.  The major axis of component A is only ~1.62″ (100 pc), while that of component B is only ~0.78″ (50 pc).  The faint plume of stars to the northwest  of A and B seen in the ground-based images of HS 0822+3542 (Kniazev et al. 2000, P03 and Gil de Paz et al. 2003) is detected in all of the images after smoothing to a  resolution comparable to the ground-based images of the object, ~0.6″ pix$^{-1}$.  The maximum extent of the object including the faint plume is ~400 pc, making it comparable in size to POX 186 (Corbin & Vacca 2002).

*2.2  Photometry and Comparison with Population Synthesis Models*

We performed photometry on components A and B using the IRAF task "polyphot," fitting polygonal apertures around them, setting a sky annulus approximately 6″ away from their centers, and converting from count rates to fluxes using the conversion factors in the image headers.  We additionally performed photometry on the brighter point sources in the images, numbered on Figure 1.  The resulting AB magnitudes and colors are presented in Table 1, along with the associated absolute magnitudes in the F550M band.  The absolute magnitudes of the point sources all fall in the range of supergiant stars under the assumption that they lie at the adopted distance to the galaxy of 12.7 Mpc.  The flux in the starburst region at the center of component A corresponds to an equivalent of ~55 OB stars, which appear to be collected into a central bright cluster with ~5 – 6 smaller associations nearby (Figure 1).  The central cluster thus falls short of being a "super" star cluster under the criterion of having a mass ~$10^4$ M$_\odot$  (see

Whitmore 2003 and references therein). The brighter point sources in region B may represent individual stars, or small OB clusters or associations.

Figure 2 shows the fluxes of components A and B as measured in all three filters. The combined F550M filter fluxes of components A and B agrees with the flux level of the spectrum presented by P03 at the corresponding wavelength. We attempted to fit these fluxes using Bruzual & Charlot (2003) population synthesis models. The models chosen represent a simple stellar population (instantaneous burst), a Chabrier (2003) initial mass function, and a metalicity of Z = 0.0004. This metalicity corresponds to approximately 1/48 $Z_\odot$, which is the closest in the Bruzual & Charlot (2003) library to that measured for HS 0822+3542. We find that the fluxes cannot be fitted by a single burst model of any age, nor by a model of continuous star formation. They can however be fitted to within errors by the addition of a young (few Myr old) component and an underlying older (several Gyr old) component. Figure 2 shows the result for the addition of the 1 Myr and 10 Gyr models, which fits the measured fluxes well ($\chi^2 \cong$ 1). More precise constraints on the ages of the respective components requires additional optical and near infrared fluxes of each component, optimally provided by spectra. In particular, there may be a contribution from a population of intermediate age (~1 Gyr) stars, as has been determined from analyses of the spectra of blue compact dwarf galaxies (Kong et al. 2003; Westera et al. 2004).

The F330W, F550M, and F814W filters exclude the stronger emission lines in the spectrum of the object. However, the nebular continuum emission at these wavelengths can be significant during the first few Myr of a starburst (Leitherer & Heckman 1995). To address this question we estimated the flux of the combined H and He nebular continuum surrounding the starburst region of component A near the central wavelengths of the filters. Using the emission-line fluxes measured from the Sloan Digital Sky Survey spectrum, we find nebular temperatures and densities of T $\cong$ 20,000° K and $N_e \cong$ 390 cm$^{-3}$. The specific values of the combined H and He nebular continuum fluxes we obtain assuming 55 O7V stars as estimated above are plotted with crosses on Figure 2. Comparing these values with the fluxes measured in each filter shows that the combined H and He nebular continuum emission may contribute significantly (~10% – 20%) to the total flux in each filter, but does not dominate it. In particular, it does not seem possible to produce the flux measured in the F814W filter without a significant population of evolved stars. We thus conclude that the emission measured in each filter and seen in Figure 1 is mainly stellar.

There is a large difference between  the surface brightnesses of components A and B and the surrounding faint plume of stars.  We measure a surface brightness of approximately 23 mag arcsec$^{-1}$ in the F814W filter, where the plume is brightest, after subtracting off the background level measured from a region away from the plume.  By contrast, the surface brightness of component B and the ring of component A is approximately 20.6 mag arcsec$^{-1}$.  This factor of ~100 difference in surface brightness further indicates that A and B are physically distinct objects, as opposed to star clusters within a single galaxy as discussed by P03.

## 3.  DISCUSSION

The small sizes of components A and B is remarkable, lying between globular clusters and objects previously classified as galaxies.  While these components are comparable in size to ultracompact dwarf galaxies found in galaxy clusters (see Drinkwater et al. 2003), they differ in that they reside in a void, are actively star-forming, and do not appear to be the remnants or fragments of larger galaxies.  HS 0822+3542 is morphologically similar to the blue compact dwarf galaxy POX 4, which also consists of a ring-like primary component and a smaller secondary component (Mendez & Esteban 1999).  These authors interpret the ring and starburst of the primary component of POX 4 to be the result of a penetrating collision by the secondary component, which is also appears to be a plausible explanation of HS 0822+3542 (see, e.g., Athanassoula, Puerari & Bosma 1997 and references therein).

The absolute magnitudes and colors of the components correspond to $B$-band luminosities of approximately $5 \times 10^6$ L$_\odot$ (component A) and $9 \times 10^5$ L$_\odot$ (component B).  Thuan (1987) estimates the ratio of mass to $B$-band luminosity of blue compact dwarf galaxies to be $2 - 4$, based on dynamical masses.  Using these ratios yields masses of $(1 - 2) \times 10^7$ M$_\odot$ (component A) and $(2 - 4) \times 10^6$ M$_\odot$ (component B).  Combining their projected separation of approximately 80 pc with these masses yields a free-fall time for the components  of ~$2 - 5$ Myr.  This is short enough for a collision between them to have triggered their starbursts.  Interaction between A and B before their recent collision could also have created the surrounding plume of stars through tidal effects, likely with B being tidally stripped by A.  In addition, the evidence that the point sources identified around A and B are supergiant stars at the same distance indicates that stars were forming in the plume as recently as $\sim 10^7$ years ago, which is consistent with the dynamical timescale of the A/B system based on the extent of the plume and the above mass estimates.

This interpretation differs from that offered by P03, who argue that HS 0822+3542's star formation was triggered by interaction with SAO 0822+3545 between $\sim$0.2-0.3 Gyr ago.  Such an interaction may have occurred, and may have contributed to the formation of the stellar plume, but the interaction of A and B seems likely to dominate the recent star formation of the system.  It seems likely that Components A and B along with the surrounding stars in the plume will eventually coalesce into a single very low-mass and quiescent dwarf elliptical galaxy.  A similar interpretation was offered by Corbin & Vacca (2002) for POX 186, which has remarkably similar properties to HS 0822+3542, including in size, mass and environment.

The evidence that components A and B contain a population of stars $\sim$10 Gyr old (Figure 2) is consistent with the general finding that blue compact dwarf galaxies contain both young and old stars  (e.g. Raimann et al. 2000; Kong et al. 2003; Westera et al. 2004).  This finding is of particular interest in view of recent evidence that the prototypical extremely metal-poor galaxy I Zw 18 contains no stars older than $\sim$ 1 Gyr  (Izotov & Thuan 2004; Östlin & Mouhcine 2005).  This suggests that the similar metal deficiency of HS 0822+3542 is the result of its low mass and the consequent loss of chemically enriched gas via supernova ejection (see MacLow & Ferrara 1999; Martin, Kobulnicky & Heckman 2002), as opposed to youth.  Extreme metal deficiency among blue compact dwarf galaxies may thus not be a unique indicator of youth.  However, given the subgalactic size of components A and B and the likelihood of their ultimate coalescence, it may be fair to describe the HS 0822+3542 system as being a dwarf galaxy in a stage of assembly from smaller components.  The masses we estimate for A and B are close to the minimum mass scale for galaxies forming via atomic cooling within cold dark matter halos (White & Rees 1978), and the formation of dwarf galaxies in voids was predicted by early cold dark matter models of galaxy formation (Dekel & Silk 1986).  We also note that the H I mass of HS 0822+3542 is approximately $1.3 \times 10^7 \, M_{\odot}$ (Chengalur et al., in preparation, as quoted by P03) which in comparison with the estimated masses of components A and B yields a ratio of gas mass to stellar mass ~1. This suggests that the object is still converting much of its gas to stars. Combining the estimate of the object's star formation rate from Kniazev et al. (2000) with our mass estimate yields a high specific or mass-normalized star formation rate (see Kauffmann et al. 2004 and references therein) of $\log(SFR/M_\odot) \cong$ -9.2 (where SFR is in units of $M_\odot$ $yr^{-1}$), consistent with this interpretation.

This work was supported by NASA through grant G0-10180.06-A to Arizona State University from the Space Telescope Science Institute. The Space Telescope Science Institute is operated by the Association of Universities for Research in Astronomy, Inc. under NASA contract NAS 5-26555.

REFERENCES

Athanassoula, E., Puerari, I., & Bosma, A. 1997, MNRAS, 286, 284

Bruzual, G. & Charlot, S. 2003, MNRAS, 334, 1000

Chabrier, G. 2003, PASP, 115, 763

Corbin, M.R. & Vacca, W.D. 2002, ApJ, 581, 1039

Dekel, A. & Silk, J. 1986, ApJ, 303, 39

Drinkwater, M.J., Gregg, M.D., Hilker, M., Bekki, K., Couch, W.J., Ferguson, H.C., Jones, J.B.,

    Phillipps, S. 2003, Nature, 423, 519

Gil de Paz, A., Madore, B.F. & Pevunova, O. 2003, ApJS, 147, 29

Izotov, Y.I. & Thuan, T.X. 2004, ApJ, 616, 768

Kauffmann, G. et al. 2004, MNRAS, 353, 713

Kniazev, A.Y. et al. 2000, A&A, 357, 101

Kniazev, A.Y., Grebel, E.K., Hao, L., Strauss, M.A., Brinkmann, J. & Fukugita, M. 2003, ApJ, 593, L73

Kong, X., Charlot, S., Weiss, A., and Cheng, F.Z. 2003, A&A, 403, 877

Leitherer, C. & Heckman, T.M. 1995, ApJS, 96, 9

MacLow, M. & Ferrara, A. 1999, ApJ, 513, 142

Martin, C., L., Kobulnicky, H.A. & Heckman, T.M. 2002, ApJ, 574, 663

Mendez, D.I. & Esteban, C. 1999, AJ, 118, 2723

Östlin, G. & Mouhcine, M. 2005, A&A, 433, 797

Pustilnik, S.A., Kniazev, A.Y., Pramskij, A.G., Ugryumov, A.V. & Masegosa, J. 2003, A&A, 409, 917

(P03)

Raimann, D., Bica, E. Storchi-Bergmann, T., Melnick, J. & Schmitt, H. 2000, MNRAS, 314, 295

Schlegel, D.J., Finkbeiner, D.P. & Davis, M. 1998, ApJ, 500, 525

Thuan, T.X. 1987, in Nearly Normal Galaxies, ed. S.M. Faber (New York: Springer), 67

White, S.D.M. & Rees, M.J. 1978, MNRAS, 183, 341

Whitmore, B. 2003, in Extragalactic Globular Cluster Systems, Proceedings of the ESO

Workshop held in Garching, Germany, ed. M. Kissler-Patig, 336

Westera, P., Cuisinier, F., Telles, E. & Kehrig, C. 2004, A&A, 423, 133

TABLE 1

AB MAGNITUDES AND COLORS[b]

| Object | F550M | F330W – F550M | F550M – F814W | $M_{F550M}$ |
|--------|-------|---------------|---------------|-------------|
| A | 19.55 ± 0.08 | -0.46 ± 0.16 | -0.07 ± 0.02 | -11.02 ± 0.08 |
| B | 21.37 ± 0.18 | -0.47 ± 0.27 | -0.07 ± 0.04 | -9.15 ± 0.18 |
| 1 | 24.3 ± 0.7 | 1.2 ± 2.2 | -0.4 ± 1.3 | -6.2 ± 0.7 |
| 2 | 24.3 ± 0.7 | 0.5 ± 1.5 | -0.3 ± 1.2 | -6.2 ± 0.7 |
| 3 | 24.3 ± 0.7 | >1.8 ± 0.7 | 0.8 ± 1.0 | -6.2 ± 0.7 |
| 4 | 25.0 ± 1.0 | -0.5 ± 1.9 | -0.4 ± 1.5 | -5.5 ± 1.0 |
| 5 | 24.7 ± 0.9 | >1.5 ± 0.9 | 1.0 ± 1.2 | -5.8 ± 0.9 |
| 6 | 24.2 ± 0.7 | >1.9 ± 0.7 | -0.3 ± 1.2 | -6.3 ± 0.7 |

[b]$M_{F550M}$ is calculated for a distance of 12.7 Mpc (§ 1).

FIGURE CAPTIONS

 FIGURE 1. -  Color image of HS 0822+3542 created from the ACS/HRC F330W, F550M and F814W images, and displayed on a logarithmic intensity scale.  The object labeled "BG" is a probable background galaxy.  The objects labeled 1-6 are point sources.  The inset shows a gray-scale version of the color image in the region of components A and B to better show their internal structure.

FIGURE 2. - Comparison of the F330W, F550M and F814W fluxes of components A with B with Bruzual & Charlot (2003) instantaneous burst stellar population synthesis models.  The models assume a Chabrier (2003) initial mass function, and a metalicity of $Z = 0.0004$.  The dashed line represents the addition of the 1 Myr model and the 10 Gyr model.   The crosses shown in the plot for component A represent the estimated combined H and He nebular continuum flux (§ 2.2).

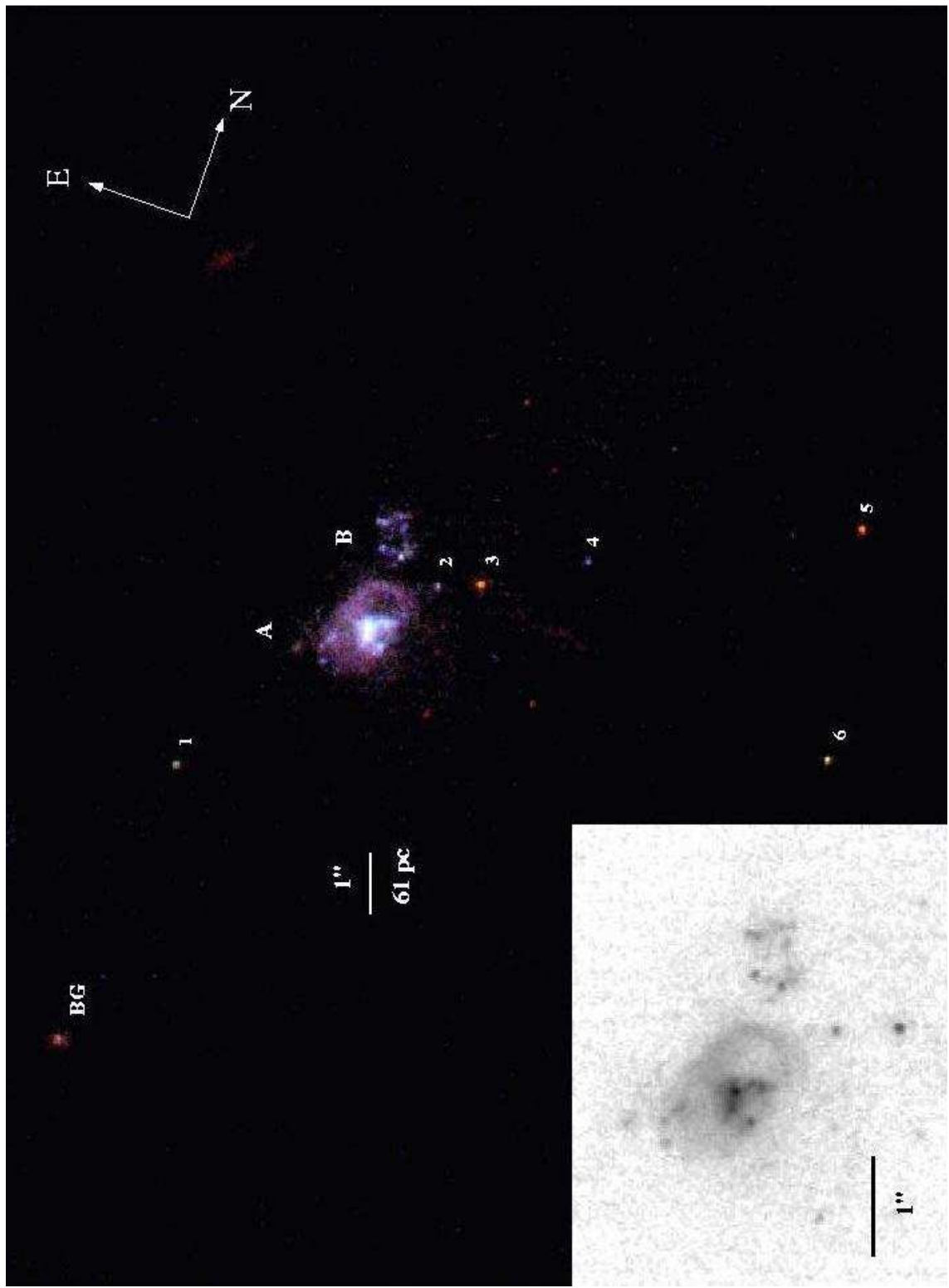

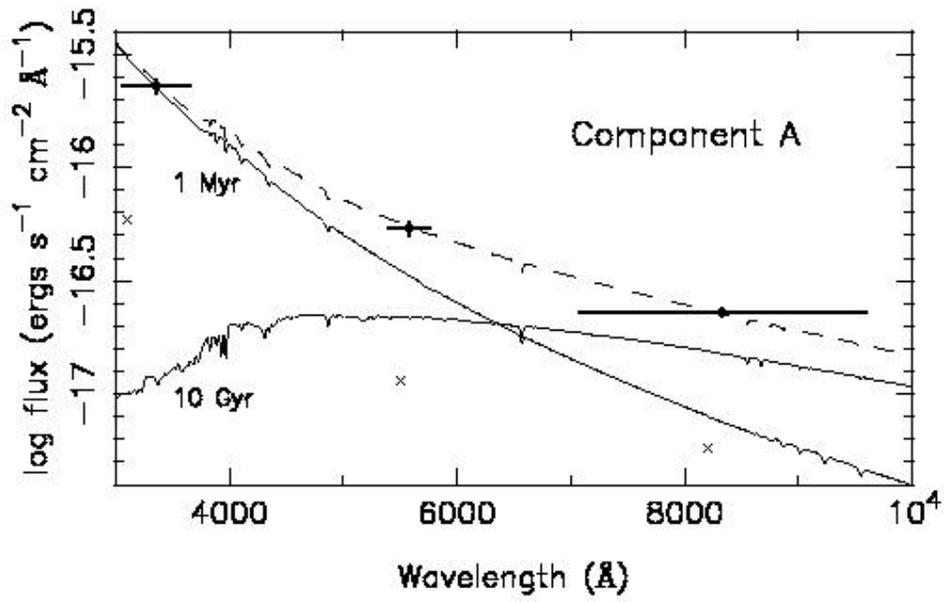

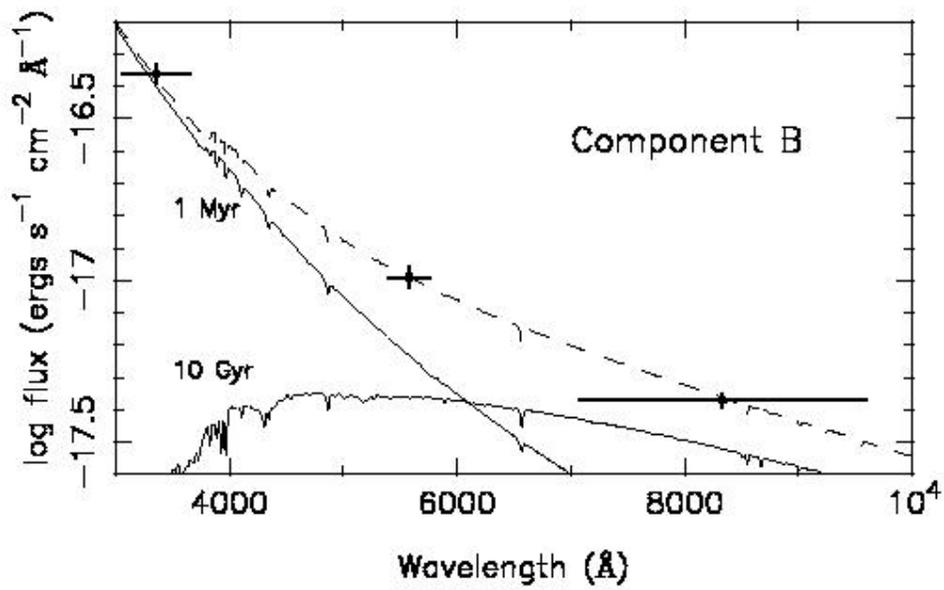